\documentclass[aps,prb,twocolumn,noshowpacs,superscriptaddress]{revtex4-2}
\usepackage{amsmath}
\usepackage{bm}
\usepackage{cases}
\usepackage{graphicx}
\usepackage{amsbsy}
\usepackage{amssymb}
\usepackage{amsmath}
\usepackage{siunitx}
\usepackage{latexsym}
\usepackage{wasysym}
\usepackage{pifont}
\usepackage{mathrsfs}
\usepackage{natbib}
\usepackage{float}
\usepackage{epstopdf}
\newfloat{widefig}{thp}{lop}
\usepackage{comment}
\usepackage{verbatim}
\usepackage{multirow,array}
\usepackage{hyperref}
\usepackage{braket}
\usepackage{xcolor}
\usepackage{listings}
\usepackage{xcolor}
\usepackage{tabularx}

\DeclareSymbolFont{epsilon}{OML}{cmm}{m}{it}
\DeclareMathSymbol{\epsilon}{\mathord}{epsilon}{"0F}

\definecolor{codegreen}{rgb}{0,0.6,0}
\definecolor{codegray}{rgb}{0.5,0.5,0.5}
\definecolor{codepurple}{rgb}{0.58,0,0.82}
\definecolor{backcolour}{rgb}{0.95,0.95,0.92}

\lstdefinestyle{mystyle}{
    backgroundcolor=\color{backcolour},   
    commentstyle=\color{codegreen},
    keywordstyle=\color{magenta},
    numberstyle=\tiny\color{codegray},
    stringstyle=\color{codepurple},
    basicstyle=\ttfamily\footnotesize,
    breakatwhitespace=false,         
    breaklines=true,                 
    captionpos=b,                    
    keepspaces=true,                 
    numbers=left,                    
    numbersep=5pt,                  
    showspaces=false,                
    showstringspaces=false,
    showtabs=false,                  
    tabsize=2
}
\begin{document}

\title{A high-performance GPU implementation of the electron-phonon Wannier interpolation and the related transport properties}
\author{Zhe Liu}
\affiliation{Institute for Advanced Study, Shenzhen University, Shenzhen 518060, China}
\author{Bo Zhang}
\affiliation{Institute for Advanced Study, Shenzhen University, Shenzhen 518060, China}
\author{Zheyong Fan}
\affiliation{College of Physical Science and Technology, Bohai University, Jinzhou 121013, China}
\author{Wu Li}
\email{wu.li.phys2011@gmail.com}
\affiliation{Institute for Advanced Study, Shenzhen University, Shenzhen 518060, China}


\begin{abstract}
The electron-phonon Wannier interpolation (EPWI) method is an efficient way to compute the properties of electron-phonon interactions (EPIs) accurately. This study presents a GPU-accelerated implementation of the EPWI method for computing transport properties, followed by a performance analysis. The implementation is tested on common systems such as aluminum and silicon. The results show complete consistency with those obtained through CPU computations. The proposed algorithm has the capability of computing the conductivity of aluminum in 20 minutes on a single NVIDIA Tesla V100 GPU, adopting a $200^3$ electron and phonon sampling grid. This speed is 173 times higher than the CPU-based algorithm, running on two nodes of the Intel Xeon Platinum 8260 CPU. Such impressive acceleration is achieved by carefully designing the algorithm to exploit the GPU's specific features. Furthermore, this methodology establishes a generic foundation for EPWI algorithms, which can be applied to other EPI-related properties.
\end{abstract}
\maketitle
\date{\today}
\section{Introduction}\label{intro}
In crystals, electron-phonon interactions (EPIs) give rise to a wide variety of phenomena, making it a critical area of interest for scientists. For example, EPIs either determine or strongly influence numerous transport properties, including the carrier mobilities of semiconductors, the electric resistivity of metals, and the thermal conductivity of solids\cite{ziman2001electrons}. They also give rise to several other phenomena, such as the transition temperatures in conventional superconductors\cite{PhysRevLett.91.166803}, the temperature-dependent optical spectra of semiconductors\cite{PhysRevB.94.075125}, Raman spectroscopy\cite{PhysRevLett.68.883}, Kohn anomaly\cite{kohn1959image}, and Peierls instability\cite{peierls1955quantum}. Computational investigations on EPIs and their effects are crucial for both fundamental and practical applications. 

The electron-phonon (el-ph) matrix element plays a key role in EPIs. Given the high cost of the density functional perturbation theory (DFPT)\cite{RevModPhys.73.515} method, calculating a large number of el-ph matrix elements is unfeasible using first-principles methods. Nevertheless, to achieve accurate results, a large number of matrix elements are required. Before 2007, few studies\cite{PhysRevB.85.115317, PhysRevLett.91.166803} used first-principles calculations to explore EPIs and their effects on materials. It was not until the development of the electron-phonon Wannier interpolation (EPWI) method\cite{giustino2007electron} that the situation began to shift. The EPWI approach employs a state-of-the-art maximally localized Wannier functions (MLWF) method\cite{marzari2012maximally}, which offers a highly accurate and efficient means of interpolating el-ph matrix elements to any arbitrary point with minimal cost. With the completion of the interpolation process, various EPI-related properties of materials can be computed using the el-ph matrix elements and additional necessary quantities. Initially, the EPW software\cite{Ponce_CPC16} implemented these calculation processes, and recently, Perturbo\cite{zhou2021perturbo}, a similar program, was designed for comparable targets.

Despite utilizing the EPWI approach to interpolate el-ph matrix elements, determining related transport properties with high accuracy, particularly for systems with large unit cells, can still be challenging. It necessitates a considerable number of sampling points to reach converged results, and the computational costs are typically positively related to both the quantity of sampling points as well as the number of atoms and Wannier functions in the unit cell. For instance, to get converged mobility of silicon at 300 K, an ultra-dense grid from $100^3$ to $300^3$ is necessary for both electron and phonon sampling\cite{Ma2018,PhysRevResearch.3.043022,PhysRevB.97.121201}. At low temperatures, the convergence of resistivities and mobilities is much harder to reach and we need even denser grids and smaller broadenings of Gaussian functions\cite{C8TC05269G_Liu}. 

In recent years, graphics processing units (GPUs) have been developing at a high speed. Due to its excellent performance on floating-point computing, GPUs have been widely employed to accelerate scientific programs and show remarkable results\cite{fan2017efficient, jia2017gpu}. Despite the progress made, a significant gap still exists in the availability of GPU-accelerated programs for the EPWI and related properties that can provide considerable speed improvements. This gap is mainly because of the difficulty to leverage the potential of GPUs. Achieving high performance on a CPU-GPU hybrid hardware structure requires a careful design of  algorithms. Fortunately, there are two conveniences for the GPU-acceleration of EPWI method. Firstly, the primary calculations of EPWI are linear tensor calculation and matrix diagonalization, which are excelling in GPU performance. Secondly, EPWI at different electron and phonon wavevectors are mutually independent, making the cost of communication negligible. Given that communication can significantly slow down the GPU performance, the GPU-acceleration of the EPWI method can be highly parallelized on GPUs with low communication costs. Therefore, the EPWI method has great potential for significant acceleration through GPUs.

The computation of EPI-related transport properties is among the most important applications of the EPWI method. By solving the Boltzmann transport equation (BTE) using el-ph matrix elements, carrier mobility, electrical conductivity, electronic thermal conductivity, and Seebeck coefficients of crystal materials can be acquired. Since this type of computation mainly consists of floating-point operations with low communication overhead, GPU acceleration could significantly enhance its efficiency. Therefore, it is of great practical value to develop a GPU-accelerated algorithm for the EPWI method, embedded in a BTE solver (EPWI+BTE). A comprehensive study of the algorithm's performance could also identify the key factors for GPU acceleration in this case. Furthermore, this methodology can also establish a generic foundation for EPWI algorithms, which can be applied to other EPI-related properties.

In this paper, we present a high-performance GPU algorithm of EPWI embedding in a BTE solver. The GPU algorithm is refactored entirely from a CPU algorithm to fit the particular features of GPUs. We discuss the structure of the algorithm and the specific details essential for maximizing the advantages of GPU. To demonstrate the performance of our algorithm, we tested it on silicon, aluminum, and other commonly used systems. All of these tests show good performance. Additionally, we compared our algorithm's performance with that of popular CPU programs designed for comparable targets, and found that our implementation exhibits superior computational performance.

The rest of this paper is organized as follows: Section \ref{Sec:Theory} provides the theoretical formulations necessary for the computation of transport properties related to EPIs. Then, Section \ref{Sec:Program} introduces the structure and technical details of the GPU algorithm. In Section \ref{Sec:Perf}, we present the performance of the algorithm, discuss the factors that affect its performance, and compare it with popular CPU programs for similar targets. Finally, we summarize the main findings in Section \ref{Sec:Conclusion}.

\section{Theoretical Formulations}\label{Sec:Theory}
\subsection{Electron-Phonon interaction}
Among all kinds of elementary excitation in pure crystals, phonons play crucial roles on the scattering of electrons at finite temperature. The perturbation Hamiltonian is represented as

\begin{eqnarray}
	\label{el-ph-Ham}
	\hat V_{e-ph}=\sum_{\nu,m,\mathbf q,n,\mathbf k}g_{mn}^\nu(\mathbf k,\mathbf q)
C_{m\mathbf {k+q}}^\dag C_{n{\mathbf k}}(\hat a_{\nu{\mathbf q}}+\hat a_{\nu\mathbf {-q}}^\dag),
\end{eqnarray}
where $\mathbf k$ and $\mathbf q$ are crystal momenta of electron and phonon, respectively; $m$ and $n$ denote band indices; $\nu$ means the mode index of phonon. 
$C_{m\mathbf {k+q}}^\dag$ and $C_{n{\mathbf k}}$ ($\hat a_{\nu\mathbf {-q}}^\dag$ and $\hat a_{\nu{\mathbf q}}$) are the creation and destruction operators of electrons (phonons). The electron-phonon matrix elements $g_{mn}^\nu(\mathbf k,\mathbf q)$ determine the coupling strength between electrons and phonons, which is formulated as
\begin{eqnarray}
	\label{el-ph}
	g_{mn}^\nu(\mathbf k,\mathbf q)=\sum_{\nu,m,\bm q,n,\bm k}\langle \varphi_{m{\bm k+\bm q}}|\partial_{\nu\mathbf q} V_{scf}|\varphi_{n{\bm k}}\rangle,
\end{eqnarray}
with $\varphi_{n{\bm k}}$ being wavefunction of electrons and $\partial_{\nu\mathbf q} V_{scf}$ being the derivative of self-consistent potential $V$ with respect to the vibration generated by an arbitrary phonon state $\nu \mathbf q$.

\subsection{Wannier Interpolation}
The Wannier interpolation method of electronic states have been widely used. It can be briefly summarized as follows. 

The Wannier functions are defined as 
\begin{eqnarray}
	|\mathbf{R} n\rangle=\frac{V}{(2 \pi)^3} \int_{\mathrm{BZ}} d \mathbf{k} e^{-i \mathbf{k} \cdot \mathbf{R}} \sum_{m} U_{m n}^{(\mathbf{k})}\left|\psi_{m \mathbf{k}}\right\rangle, 
\end{eqnarray}
where $V$ is the real-space primitive cell volume, $\mathbf R$ are lattice vectors and $U_{m n}^{(\mathbf{k})}$ are unitary matrices. By selecting specific $U_{m n}^{(\mathbf{k})}$, one can get the MLWFs, which are very useful to interpolate wavefunctions into any points with a rather coarse sample mesh by 


\begin{eqnarray}
	\sum_{n'm'}U_{nn'}^{(\mathbf{k})}\left\langle\psi_{\mathbf{k} n}|H| \psi_{\mathbf{k} m}\right\rangle U_{m'm}^{(\mathbf{k})*}=\sum_{\mathbf{R}} e^{i \mathbf{k} \cdot \mathbf{R}}\langle\mathbf{0} n'|H| \mathbf{R} m'\rangle.
\end{eqnarray}

As the Hamiltonian is diagonalized in any $\mathbf k$ subspace, the $U^{(\mathbf{k})}$ can be calculated by diagonalization procedures. Then we can easily obtain wavefunctions. 

Similarly, the dynamical matrix $D(\mathbf q)$ can also be interpolated from second order force constant $\Phi(\mathbf R)$ via 
\begin{eqnarray}
	D_{\nu\mu}(\mathbf{q}) = \sum_{\mathbf R}\frac{1}{\sqrt{m_\nu m_\mu}} \Phi_{\nu\mu}(\mathbf R) \exp(\mathrm{i}\mathbf {q}\cdot\mathbf R) ,
\end{eqnarray}
where, $\nu$($\mu$) denotes the freedom of atomic variation in a unit cell. 

Then, the phonon frequencies $\omega_{\nu \mathbf q}$ and eigenvectors $e^\nu_{i}(\mathbf q)$ can be obtained by the diagonalization of $D(\mathbf q)$
\begin{eqnarray}
	D_{ij\alpha\beta}(\mathbf{q})e^\nu_{i}(\mathbf q) = \omega_{\nu \mathbf q}^2 e^\nu_{i}(\mathbf q).
\end{eqnarray}

The Wannier interpolation of el-ph matrix element $g$ can be expressed as follow,
\begin{eqnarray}
	\begin{aligned}
		g_{m n}^{\nu}(\mathbf{k}, \mathbf{q})= \sum_{i, j, \mu, \mathbf{R}_e, \mathbf{R}_p } & U_{m i}(\mathbf{k}+\mathbf{q}) U_{j n}^{\dagger}(\mathbf{k}) \\
		& \times e^{i\left(\mathbf{k} \cdot \mathbf{R}_e+\mathbf{q} \cdot \mathbf{R}_p\right)} g_{i j}^{\mu}\left(\mathbf{R}_e, \mathbf{R}_p\right)u_{\mu\nu}(\mathbf q),
	\end{aligned}
\end{eqnarray}
where $u_{\mu\nu}(\mathbf{q})=\left(\hbar / 2 m_\mu \omega_{\mathbf{q} \nu}\right)^{1 / 2} e^{\nu}_{\mu}(\mathbf{q})$, and 
\begin{eqnarray}
\begin{aligned}
	g_{i j}^{\mu}\left(\mathbf{R}_e, \mathbf{R}_p\right)= \frac{1}{N_e N_p}\sum_{i, j, \mu, \mathbf{k}, \mathbf{q} } & U_{i m}^{\dagger}(\mathbf{k}+\mathbf{q}) U_{n j}(\mathbf{k}) \\
	& \times e^{-i\left(\mathbf{k} \cdot \mathbf{R}_e+\mathbf{q} \cdot \mathbf{R}_p\right)} g_{m n}^{\nu}(\mathbf{k}, \mathbf{q})u_{\nu\mu}^{-1}(\mathbf q).
\end{aligned}
\end{eqnarray} 

As long as $g_{i j}^{\mu}\left(\mathbf{R}_e, \mathbf{R}_p\right)$ decays rapidly with the increasing of 
$\mathbf{R}_e$ and $\mathbf{R}_p$, just like $H(\mathbf{R}_e)$ and $\Phi(\mathbf{R}_p)$, the electron-phonon interpolation method could show a rather good performance. 

It should also be noted that the mesh needed is always ultra dense, and the interpolation on different $\bm k$ and $\bm q$ points are independent with each other. Given these consideration, the EPWI method is well suited for GPU parallelization.

\subsection{Boltzmann transport equation and transport coefficients}
To get the transport properties of materials, the semi-classical Boltzmann transport equation (BTE) should be solved. The general  form of BTE can be expressed as follows. 

\begin{eqnarray}
	\frac{\partial f_{n\mathbf k}}{\partial t} = 
	\frac{\partial f_{n\mathbf k}}{\partial t}\Big|_{\text{drift}} + 
	\frac{\partial f_{n\mathbf k}}{\partial t}\Big|_{\text{scatt}} = 0,
\end{eqnarray}
where $f_{n\mathbf k}$ means the distribution on $n\mathbf k$ electronic state. The ``drift'' term denotes the variation of $f$ caused by external electric field, while the ``scatt'' term denotes that from interparticle scatterings. 

With a linear approximation along with relaxation time approximation (RTA), the scattering rate is expressed as
\begin{widetext}
\begin{equation}
	\begin{aligned}
	{\tau^{-1}_{n \mathbf{k}}}=&\frac{2 \pi}{\hbar} \sum_{\mathbf{q} p m}\left|g_{\mathbf{n k}, \mathbf{q} p}^{m \mathbf{k}+\mathbf{q}}\right|^2\left[\left(N_{\mathbf{q} p}^0+f_{m \mathbf{k}+\mathbf{q}}^0\right)\right. \delta\left(E_{n \mathbf{k}}+\hbar \omega_{\mathbf{q} p}-E_{m \mathbf{k}+\mathbf{q}}\right) \\
	&\left.\quad+\left(1+N_{-\mathbf{q} p}^0-f_{m \mathbf{k}+\mathbf{q}}^0\right) \delta\left(E_{n \mathbf{k}}-\hbar \omega_{-\mathbf{q} p}-E_{m \mathbf{k}+\mathbf{q}}\right)\right].
	\end{aligned}
\end{equation}
\end{widetext}

Then, the mean free paths (MFPs) $\mathbf{F}^0_{n \mathbf{k}} = \mathbf{v}_{n \mathbf{k}}\tau_{n \mathbf{k}}$ are computed as the start of iteration equation,

\begin{equation}
	\label{Eqn:iter}
	\mathbf{F}_{n \mathbf{k}}^{i+1}=\mathbf{F}_{n \mathbf{k}}^{0}+\tau_{n \mathbf{k}} \sum_{\mathbf{q} p m}\left(\Gamma_{n \mathbf{k}, \mathbf{q} p}^{m \mathbf{k}+\mathbf{q}}+\Gamma_{n \mathbf{k}}^{m \mathbf{k}+\mathbf{q},-\mathbf{q} p}\right) \mathbf{F}_{m \mathbf{k}+\mathbf{q}}^{i}.
\end{equation}

For convenience, the transition rates $\Gamma_{n \mathbf{k}, \mathbf{q} p}^{m \mathbf{k}+\mathbf{q}}$ and $\Gamma_{n \mathbf{k}}^{m \mathbf{k}+\mathbf{q},-\mathbf{q} p}$ here differed by a factor $f_{n \mathbf{k}}^0(1-f_{n \mathbf{k}}^0)$ from those in Ref.\cite{li2015electrical}.

Finally, we can calculate electrical conductivity, mobility, Seebeck coefficient, electronic thermal conductivity\cite{madsen2006boltztrap} through 
\begin{align}
	\sigma&=\frac{2 q^2}{N V k_B T} \sum_{n \mathbf{k}} \mathbf v_{n \mathbf{k}} \mathbf F_{n \mathbf{k}}\left(-\frac{\partial f_{n \mathbf{k}}^0}{\partial \varepsilon_{n \mathbf{k}}}\right),\\
	\boldsymbol{\sigma} \mathbf{S}&=\frac{2 q}{T V N_k} \sum_{n \mathbf{k}}\left(\varepsilon_{n \mathbf{k}}-\varepsilon_f\right) \mathbf{v}_{n \mathbf{k}} \mathbf{F}_{n \mathbf{k}}\left(-\frac{\partial f_{n \mathbf{k}}^0}{\partial \varepsilon_{n \mathbf{k}}}\right),\\
	\boldsymbol{\kappa}_e&=\frac{2}{T V N_k} \sum_{n \mathbf{k}}\left(\varepsilon_{n \mathbf{k}}-\varepsilon_f\right)^2 \mathbf{v}_{n \mathbf{k}} \mathbf{F}_{n \mathbf{k}}\left(-\frac{\partial f_{n \mathbf{k}}^0}{\partial \varepsilon_{n \mathbf{k}}}\right)-T \boldsymbol{\sigma} \mathbf{S}^2.
\end{align}

The mobility of semiconductors can be obtained via
\begin{equation}
\mu_{\alpha \beta}=\frac{\sigma^{\alpha \beta}}{n q},
\end{equation}
with $n$ being the carrier concentration and $q$ the charge of electron.

\subsection{The analysis of characteristics}
Computation of the (el-ph) matrix elements required for scattering rates calculation is a time-consuming process in the EPWI + BTE method. This process is characterized by the following aspects:

1. The majority of computations can be performed as a large number of matrix operations.

2. The scattering rate calculation for each electronic state is independent, thereby allowing parallelization without the need for communication.

3. The computational cost of each process can be estimated, making it possible to distribute and assign them equally to threads. This approach optimizes the utilization efficiency and minimizes computing power waste.

4. The valid terms for the scattering rate are constrained by energy and quasimomentum conservation, necessitating a reduction of the sampling space to an effective scattering subspace.

These aspects suggest that the EPWI + BTE method is highly suitable for GPU parallelization, given the first three reasons. Despite these advantages, implementing optimized code remains a substantial challenge. In the following section, we will present the detailed algorithm and the necessary modifications for compatibility with GPUs.


\section{Algorithm, Code Structure and Details}\label{Sec:Program}

\subsection{CPU algorithm}
\begin{figure*}[!htb]
	\includegraphics[width=13.5cm]{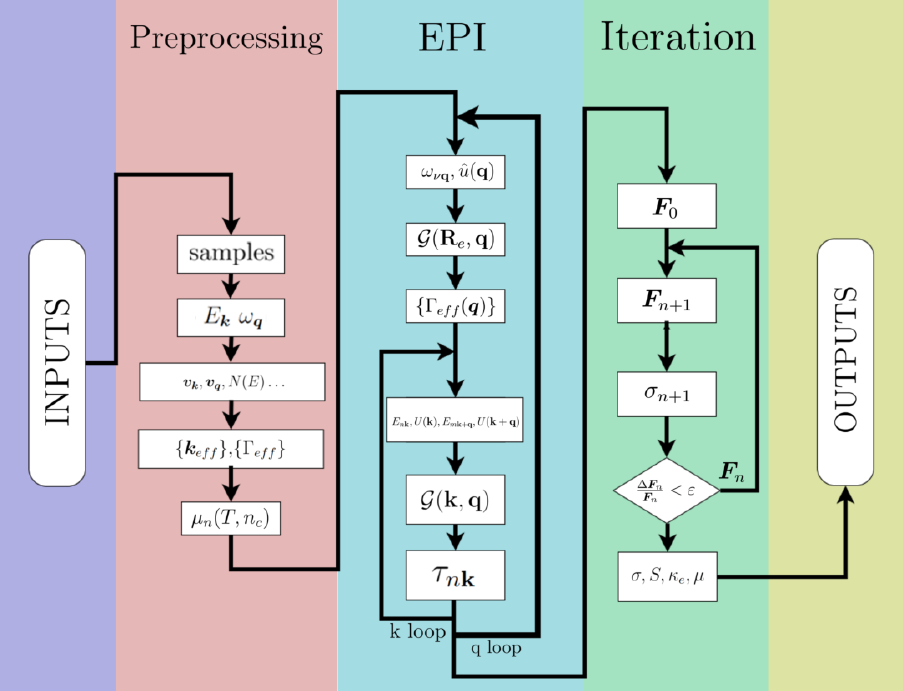}
	\caption{The flow chart of the CPU algorithm.}
	\label{fig:CPU_algo_0}
\end{figure*}

CPU algorithms have already been implemented by previous works. The structure of a typical one is demonstrated in fig.\ref{fig:CPU_algo_0}. In this algorithm, the computer reads information of system and parameters from input, such as density of sample mesh, temperatures, doping levels, etc. The information of system includes basic information of lattice, basic parameters used in first principles calculations. Besides, it also reads electron Wannier function, inter-atomic force constant, el-ph matrix element in Wannier basis from EPW software in Quantum Espresso package\cite{giannozzi2017advanced}. 

The algorithm consist of three parts: preprocessing, EPWI, iteration. At the beginning of preprocessing, Monkhorst-Pack $\mathbf k$ and $\mathbf q$ grids are generated, while taking symmetry into account. Then, Wannier interpolation procedures are employed to interpolate electron and phonon states. Following this, various quantities are computed, such as velocities, density of states, Fermi energy, etc. Additionally, the electronic states that are close to the Fermi level are selected as effective states $\{\bm k_{\text{eff}}\}$. Any quantities that are not related to el-ph scatterings can be calculated at this stage. 

One of the decisive steps of the preprocessing stage is the effective process sieve, also known as the energy conservation surface sieve. In this process, effective scattering processes are selected and counted by testing for energy and momentum conservation conditions. It is worth noting that the entire preprocessing stage can estimate the memory usage of $\Gamma$. At the end of the preprocessing stage, the chemical potential is calculated at various doping levels and temperatures.

Following the preprocessing, the EPWI, as the primary component of the process, begins. It comprises of two nested loops, which focus on data reuse and memory conservation. These loops iterate through all valid $\bm k$ states and utilize the EPWI method to compute the el-ph matrix elements for all effective scatterings, as well as the scattering rates of each electronic state. Given the large number of physical quantities required in this calculation (such as velocities, eigenvalues, and eigenvectors of millions of electron and phonon states), EPWI must be incorporated into the calculation of transport properties to alleviate the significant burden on memory. This approach is referred to as the on-the-fly method. Moreover, the order of the two loops can be swapped, and this will be discussed later. 

The EPWI process contains EPWI and the calculation of scattering rates. It costs almost all of the time of computation. Therefore, parallel acceleration is required for this part of the algorithm, which is exactly what previous CPU algorithms did. 

The message passing interface (MPI)\cite{clarke1994mpi} parallel implementation is the first to be accomplished. In this implementation, the electronic state of interest is assigned to each process. To minimize the communication overhead, a complete copy of the variables in Wannier basis, such as $g\left(\mathbf{R}_e, \mathbf{R}_p\right)$, $H(\mathbf{R}_e)$, and $\Phi(\mathbf{R}_p)$, are stored within the private memory of each process. Consequently, the scattering rates and related quantities of states in one process are computed and stored in-place, eliminating the need for communication between processes throughout the entire EPWI process. Large storage usage in the MPI implementation can be alleviated by including a shared-memory multiprocessing architecture, such as OpenMP\cite{dagum1998openmp}, into the process. With a hybrid MPI + OpenMP implementation, quantities in the Wannier basis can be shared between threads, making the memory usage more efficient.

\subsection{Overview of GPU algorithm}

The key differences between GPUs and CPUs can be summarized as follows\cite{owens2008gpu}. Firstly, GPUs are very specialised to execute a vast number of floating-point operations simultaneously, while CPUs operate sequentially. Secondly, GPUs - also known as devices here - require a call from a single CPU, also referred to as the host. The process of calling from the host incurs significant costs. Thirdly, GPUs have relatively smaller memory than CPUs, creating conflicts with the first two features. To maximize GPU utilization, it's crucial for developers to incorporate as many tasks as possible within a single device call. However, the number of tasks is limited by the maximum GPU memory. Moreover, avoiding read-write conflicts of threads and the speed of data transfer between the host and devices are additional restrictions. As a result, an effective algorithm for GPUs differs significantly from CPUs. Therefore, to run CPU algorithms on CPU + GPU hardware structures, they must be reconstructed accordingly.

The CPU-based approach divides the task into many ultra-fine pieces, including individual scattering processes, which are computed sequentially across multiple threads. However, this approach is unsuitable for GPU-based computing. Transmitting ultra-fine tasks to GPU memory and computing them one by one results in excessive processing time for data transfer between the CPU and GPU. 
To leverage the benefits of GPUs, the bulk of the data must first be located in GPU memory. Temporary intermediate data can be packaged and computed in batch forms. Therefore, the primary objective of GPU acceleration is to develop suitable data structures and workflows that enable efficient processing of batched data and computations to take full advantage of GPU acceleration.

In addition, there is another fact that is worth emphasizing. Although the EPWI process is the most time-consuming in the CPU program, focusing GPU utilization solely on this part is not practical. That's because if the main part is reconstructed and performed efficiently on the GPU, it can significantly reduce the computation time. Thus, the relative computational time of other parts, such as preprocessing, will increase noticeably. Therefore, it is essential to accelerate other parts using the GPU as well.

\begin{figure*}[!htb]
	\includegraphics[width=13.5cm]{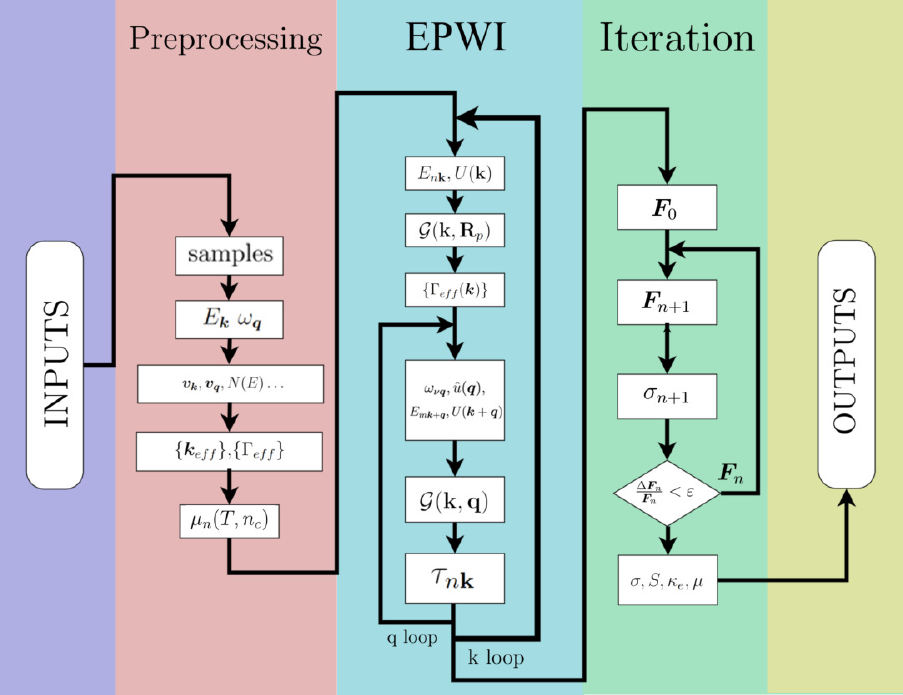}
	\caption{The flow chart of the final version of GPU algorithm.}
	\label{fig:GPU_algo_0}
\end{figure*}

To develop an algorithm well-suited for GPUs, we completely refactored the CPU algorithm. The majority of the new code was written in CUDA C++\cite{zone2020nvidia}, which was later embedded into old serial CPU codes written in Fortran. The flowchart of this algorithm is presented in fig.\ref{fig:GPU_algo_0}, where all the heavy computations are processed on the GPU.

The permutation in the calculation order may be noted. Its benefits will be analyzed in Sec.\ref{algo:loop}. Before discussing the loop ordering, the significance of data structure is discussed in Sec.\ref{algo:data}. Additionally, the implementation of the iteration process is described at the end of this section.

With a few modifications, this entire program structure can be utilized to compute other properties related to the EPIs. Specifically, the sieve and intermediate quantities in preprocessing and EPWI processes should be adjusted to new targets. Furthermore, the iteration process must be replaced by corresponding post-processing calculations. Other than these modifications, all other parts of this algorithm can be applied to new objectives.

\subsection{ Data organization and batched computing}\label{algo:data}

\subsubsection{basic principles}
A well-designed data structure is critical to the performance of GPU algorithms. GPUs execute single instruction/multiple data (SIMD) operations, which require that the data for the operations be stored contiguously and aligned on the byte boundaries of memory. Therefore, this rule should be followed when writing kernel functions that run on the GPU.

To discuss the storage methods of arrays in GPU memory, we represent an array $A$ as $A(n_1,n_2,n_3,\dots)$. This means the array is stored in column-major format with the specified index order. For instance, a zero-based rank 2 array $A_{n\times m}$ represented as $A(n,m)$ has $n$ rows, $m$ columns, and its element $A_{ij}$ is located at $(n*(j-1) + i)$ in memory. For a 3-rank array $A(n_1,n_2,n_3)$, $A_{ijk}$ is located at $(i+n_1*(j-1)+n_1*n_2*(k-1))$. 
This storage method implies that we store elements in a column contiguously in memory and this is required for coalesced memory access.

The storage of arrays also has a significant impact on the functionality of library functions. When utilizing library functions, such as cuBLAS, for matrix operations, the data must conform to specific demands. Specifically, for batched diagonalization or multiplication, the objective matrices must have the same size and be consecutively stored in GPU memory.
In the most common batched multiplication, we compute
\begin{equation*}
	C_{\alpha \gamma}(x) =\sum_\beta A_{\alpha\beta}(x) B_{\beta \gamma}(x)
\end{equation*}
using the cublasZgemmStridedBatched function, where x denotes the size of each batch, and indices $\alpha$ and $\gamma$ in matrices A and B represent all indices not involved in the summation. 
To meet the demands of the library function, $A_{\alpha\beta}(x)$ must be stored as $A(\alpha,\beta,x)$. Additionally, when the storage of $C$ is fully capable of accommodating the succeeding operation, any redundant data reorganization can be avoided. Otherwise, an inappropriate data storage mode may result in unnecessary data reconstruction between steps or divide a single process into multiple segments, causing an increase in the overhead of kernel function calls. 
To achieve the highest level of performance, it is essential to prevent such side-effects. Therefore, careful data structure design is crucial for avoiding such issues in computations and enhancing the performance of GPUs.


\subsubsection{ The data structure in the GPU algorithm}

The central equation of EPWI is given by
\begin{eqnarray}
	\begin{aligned}
		\label{eq:GPU_EPI}
		g(\mathbf{k}, \mathbf{q})=&\bigg\{ U(\mathbf{k}+\mathbf{q})\Big[\big(gP^{el}(\bm k) U^{\dagger}(\mathbf{k}) \big)P^{ph}(\bm q)\Big]\bigg\}u(\mathbf q),
	\end{aligned}
\end{eqnarray}
with 
\begin{eqnarray}
	\begin{aligned}
		P^{el}_{R_e k} =& e^{i\mathbf{k} \cdot \mathbf{R}_e},\\
		P^{ph}_{R_p q} =& e^{i\mathbf{q}\cdot\mathbf{R}_p}.\notag
	\end{aligned}
\end{eqnarray}

The brackets in Eq.\ref{eq:GPU_EPI} indicate the order of computation, consistent with the workflow in Fig.\ref{fig:GPU_algo_0}. As shown in the formula, all computations can be viewed as matrix multiplication, except for the preparation processes of the necessary data, which need involve matrix diagonalization and certain floating-point calculations. 

\begin{figure*}[!htb]
	\includegraphics[width=13.5cm]{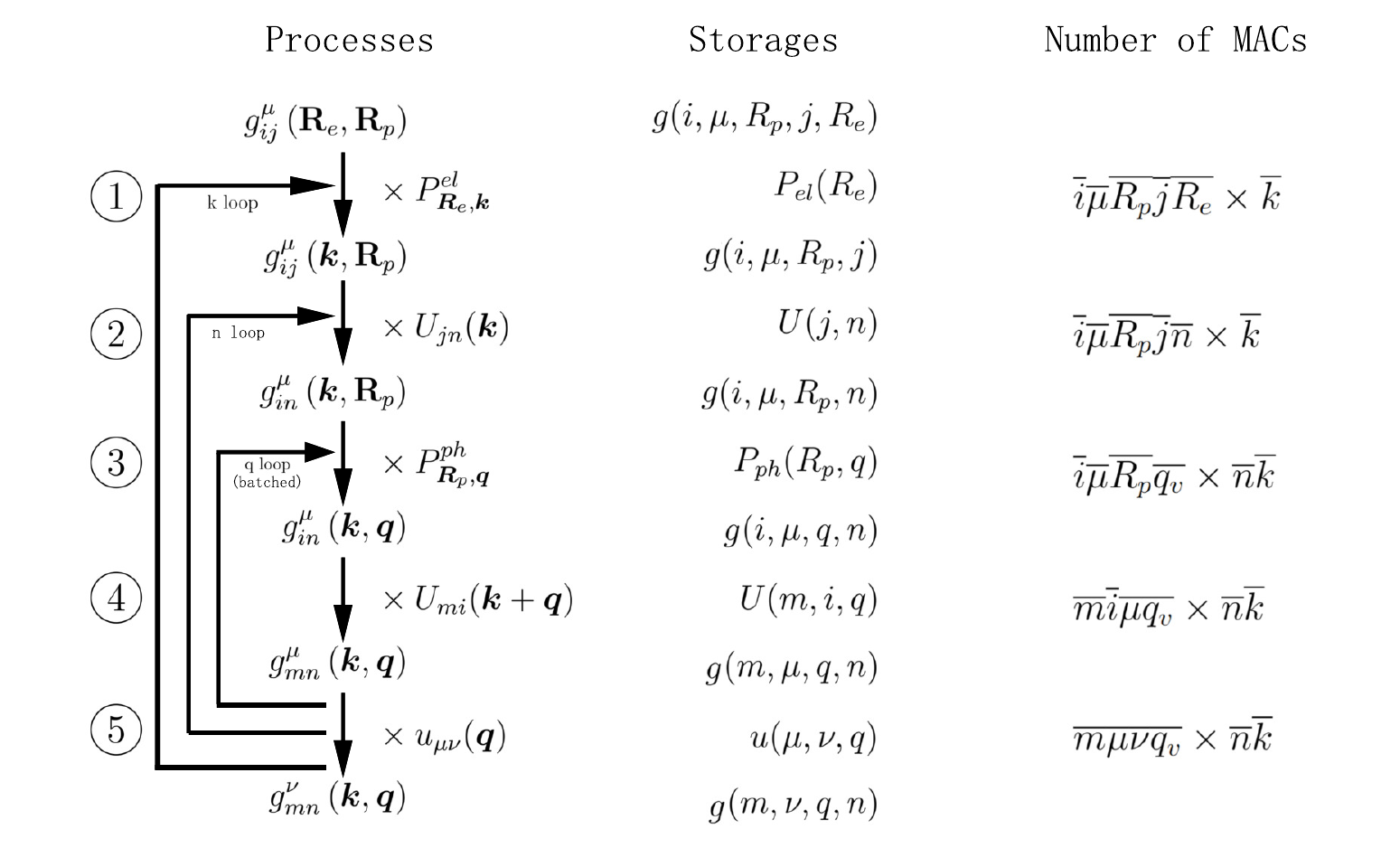}
	\caption{The flow chart, data structure, and computation volume of q-batch algorithm. The computing density is the computation volume for each time. It's the terms before ``$\times$'', while the one after ``$\times$'' denotes the number of times it is executed.}
	\label{fig:qbatdata}
\end{figure*}

The data structure shown in the second column of Fig.\ref{fig:qbatdata} was designed based on the fundamental principle and workflow outlined above. It satisfies the demands of library functions throughout the entire process and data reconstruction is minimized.

The preparation of the required matrix in Eq.\ref{eq:GPU_EPI} are computed in the batched form, by a series of GPU kernel functions, or cuSolver library functions for batched diagonalizations. Then, all matrix multiplications are conducted through functions in the cuBLAS library.

\subsection{Batch method, loop ordering and performance }\label{algo:loop}

\begin{figure*}[!htb]
	\includegraphics[width=13.5cm]{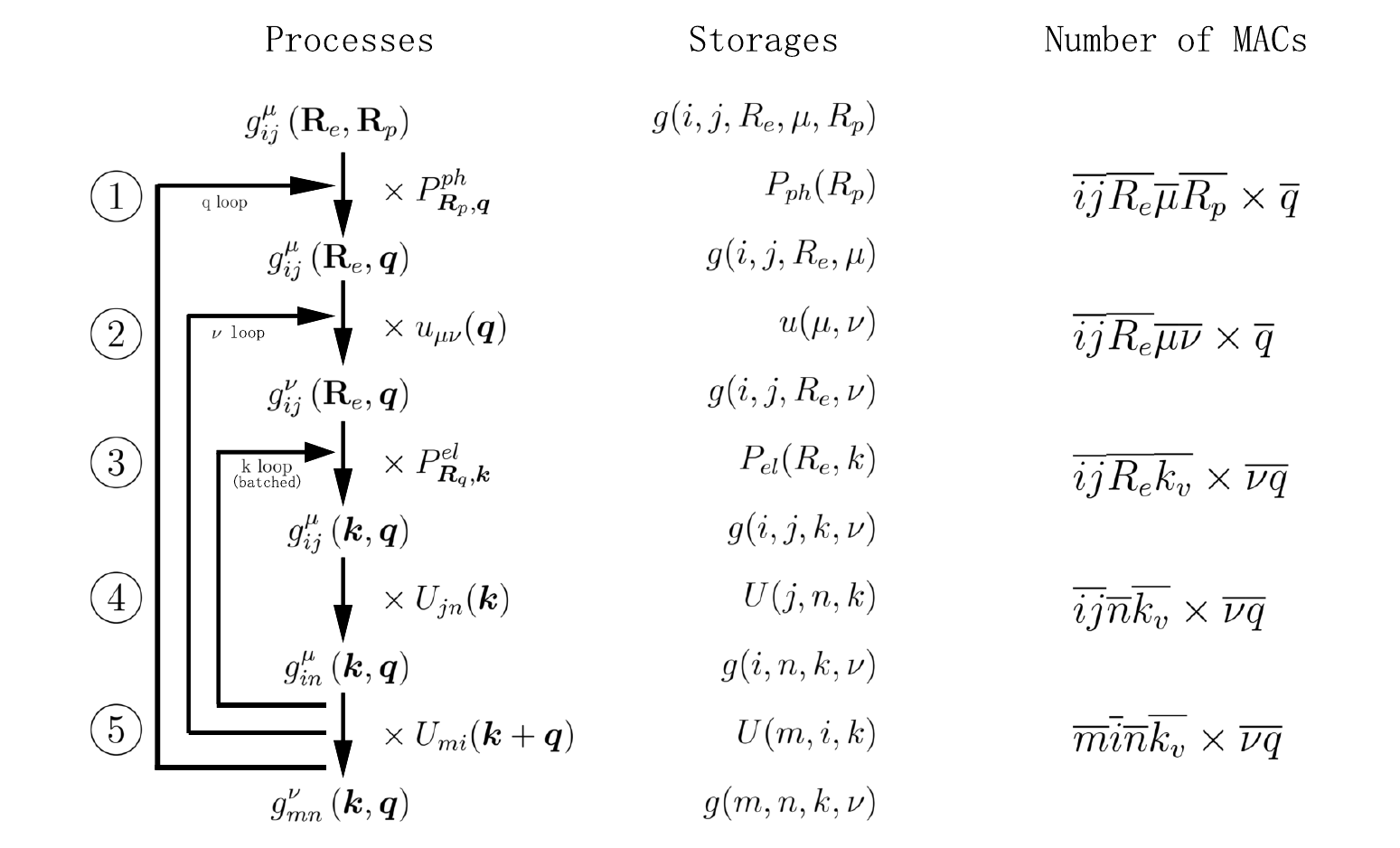}
	\caption{The flow chart, data structure, and computation volume of k-batch algorithm. The computing density is the computation volume for each time. It's the terms before ``$\times$'', while the one after ``$\times$'' denotes the number of times it is executed.}
	\label{fig:kbatdata}
\end{figure*}

The computational ordering in Eq.\ref{eq:GPU_EPI} is different from loop in CPU. The reason for this is our preference towards the q-batch parallelization method over the k-batch method. The reason is that we prefer the q-batch parallelization method over the k-batch method. A detailed analysis of the estimated computational volume and density can illustrate the benefits of the q-batch method. In Fig.\ref{fig:kbatdata}, the workflow, data structure, and computation volume of k-batch algorithm are presented for comparison.

To simplify expressions, we hereby define $\overline{X}$ as the total number of possible values for $X$. For example, $\overline{i}$ denotes the number of Wannier bands, $\overline{\nu}$ denotes the number of modes, $\overline{k_v}$ denotes the number of valid $\bm k$ points, and $\overline{\bm R_e}$ denotes the number of lattice vectors used in the sum of the electronic Wannier basis.

To obtain each necessary scattering process, diagonalization processes are required to calculate $\mathrm{u}_{\nu\mu}(\mathbf{q})$ and $U{(\mathbf{k})}$. The number of diagonalization processes required is $2\cdot\overline{kqpairs}$, which depends only on the total number of $(\bm{k},\bm{q})$ pairs associated with valid processes, denoted by ``kqpairs'', and does not change with the computational order.

With the exception of diagonalization, the majority of EPWI computation can be treated as matrix multiplication. In a matrix multiplication such as $A_{m\times n}\times B_{n\times k}=C_{m\times k}$, a fundamental operation for processors is the multiply-accumulate (MAC) operation, which is performed $n$ times for each element in $C$. The total number of operations for all elements in $C$ is $n\cdot m\cdot k$. By this approach, we can estimate the order of magnitude of operations throughout the EPWI process.

The third column of Fig.\ref{fig:qbatdata} and Fig.\ref{fig:kbatdata} presents the estimation of the computational task volume for the two algorithms in terms of the number of MAC operations. The $\overline{q_v}$ and $\overline{k_v}$ are defined as follows:
\begin{equation}\label{Eq:kqpairs}
	\overline{k_v}\overline{q} \approx \overline{q_v}\overline{k} = n_{kqpairs}.
\end{equation}
where $n_{kqpairs}$ denotes the total number of valid $(k,q)$ pairs. Here,  $\overline{q_v}$ ($\overline{k_v}$) refers to the average number of valid $q$($k$), related to a specific $k$($q$). 

In both charts, we observe that the most time-consuming step is the third step in the diagram. The computational volumes for the q-batch and k-batch modes are given by $V_{{\text{q-batch}}}=\overline{i}\overline{\mu}\overline{R_p}\overline{q_v}\times\overline{n}\overline{k}$ MACs and $V_{{\text{k-batch}}}=\overline{i}\overline{j}\overline{R_e}\overline{k_v}\times\overline{\nu}\overline{q}$ MACs, respectively. Their ratio is 
\begin{equation}
	\frac{V_{\text{q-batch}}}{V_{\text{k-batch}}}= \frac{\overline{n}}{\overline{i}}\frac{\overline{R_p}}{\overline{R_e}}\frac{\overline{q_v}\overline{k}}{\overline{k_v}\overline{q}} = \frac{\overline{n}}{\overline{i}}\frac{\overline{R_p}}{\overline{R_e}}.
\end{equation}

We note that $\frac{\overline{R_p}}{\overline{R_e}}\le 1$, since the coarse grid of $\bm q$ is always smaller or the same size as that of $\bm k$. Additionally, $\frac{\overline{n}}{\overline{i}} < 1$, since $i$ ranges over all Wannier states while $n$ is chosen only near the Fermi level. 
As a result, ${V_{\text{q-batch}}}$ is significantly smaller than ${V_{\text{k-batch}}}$. Consequently, the q-batch mode can substantially reduce the total computational volume.

However, the performance enhancement cannot be attributed solely to the differences in computational volume. The q-batch parallelization method can also make computational tasks more dense, leading to improved performance. The density of computing can be estimated from the computational volume of each loop. For example, the computational density of the third step is given by
\begin{eqnarray}
V_{q-batch}^{each-loop}=\overline{i}\overline{\mu}\overline{R_p}\overline{q_v}, \\
V_{k-batch}^{each-loop}=\overline{i}\overline{j}\overline{R_e}\overline{k_v}.
\end{eqnarray}

\begin{figure*}[htb]
	\includegraphics[width=13.5cm]{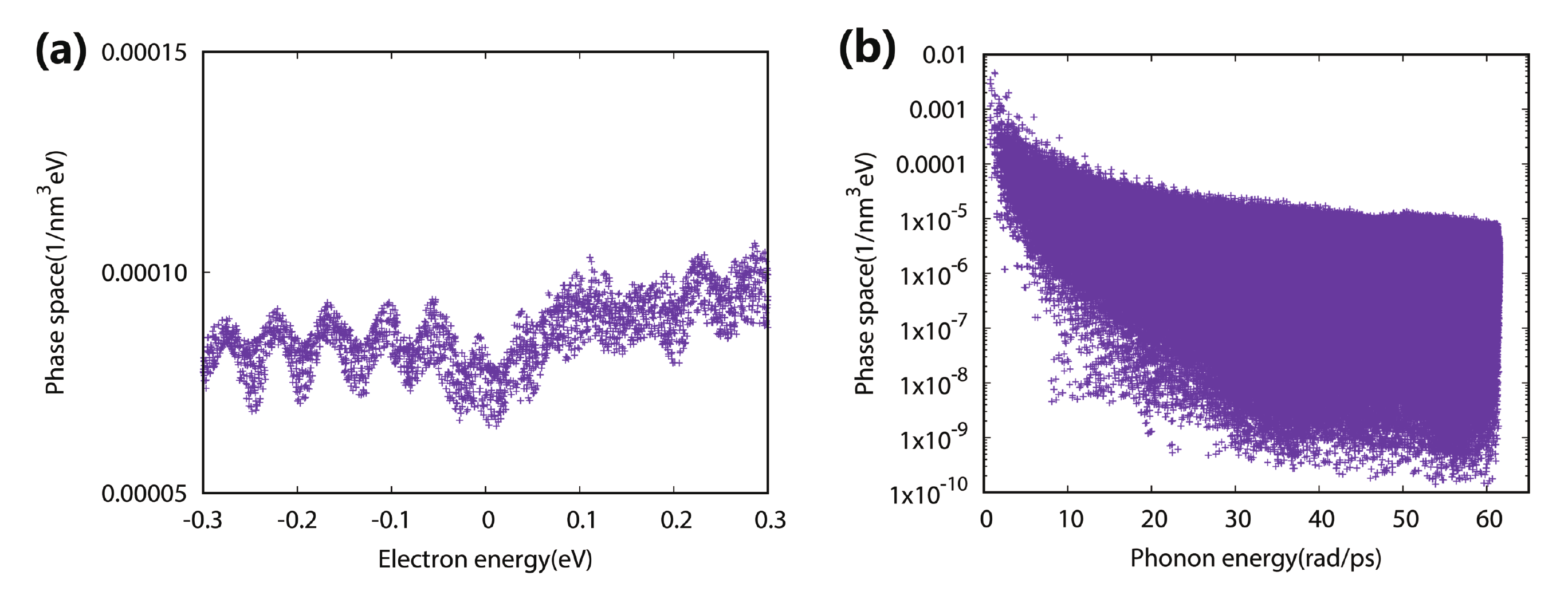}
	\caption{A two valley model is used to show the effective phase space (yellow) of (a) $\bm k=\Gamma$ (band edge) and (b) $\bm q_0$ (a common point). 
	The number of effective processes related with a specific electron (phonon) state $n_{kqpairs}(
	\bm k)$ ($n_{kqpairs}(\bm q)$) is proportional to the volume of the effective phase space.}
	\label{fig:phase_kq}
\end{figure*}

Since only a limited number of electronic states near the Fermi level are used in the calculation, while almost all phonons are used, the value of $\overline{q}$ greatly exceeds the value of $\overline{k}$. Considering Eqn.\ref{Eq:kqpairs}, it can be deduced that $\overline{q_v}$ significantly surpasses $\overline{k_v}$. This fact also suggests that the average effective phase space of a phonon is much less than that of an electron. 
A sketch of the electron and phonon phase spaces is shown in Fig.\ref{fig:phase_kq} to clearly illustrate this concept. In this example, we use a conduction band model of a semiconductor with two valleys. As shown in Fig.\ref{fig:phase_kq}(a), an electronic state at the edge of the band can be scattered to any available final states in both valleys. Consequently, it has a large effective phase space. In contrast, for a specific phonon wavevector $\bm q_0$ in Fig.\ref{fig:phase_kq}(b), only very few electron states near the band edge can be connected with this phonon wavevector. Although the phase space of electrons and phonons in metals are more complicated, the conclusion is still valid. In the Sec.\ref{result:perf}, we will present specific data on the phase space of electrons and phonons in aluminum. 

In conclusion, compared to the k-batch method, the q-batch method can significantly reduce the computation required, improving the computation density. Both effects can enhance the algorithm's performance.

\subsection{Compression storage and iteration}
As is shown in Eqn.\ref{Eqn:iter}, the transition rates $\Gamma$ are needed to perform the iteration. They are computed and stored during the EPWI process. It is important to note that the summation $\Gamma_{n,m}(k,k+q)=\sum_{p}\left(\Gamma_{n \mathbf{k}, \mathbf{q} p}^{m \mathbf{k}+\mathbf{q}}+\Gamma_{n \mathbf{k}}^{m \mathbf{k}+\mathbf{q},-\mathbf{q} p}\right)$ always appears as a whole. Therefore, we choose to store these valid summation terms instead of each $\Gamma$. Since they are sparse, it is efficient to store them in compressed sparse row (CSR) format. The detailed description of the data format is as followed. 
\begin{itemize}
	\item $k_{\text{eff}}$ stores the information of effective k points one by one. 
	\item $V_{\Gamma}$ stores the value of $\Gamma_{n,m}(k,k+q)$ with indices $(n,k)$ following the order in $k_{\text{eff}}$. 
	\item $I_{\Gamma}$ stores the index of $k+q$ points and the band index $m$ of each element in $V_{\Gamma}$. 
	\item $S_{\Gamma}$ stores the start point of each electronic state $(n,k)$ in the array $V_{\Gamma}$. 
\end{itemize}
Since all wavevectors are on a uniform grid, their integer index in the grid is stored instead of the float coordinates to reduce the memory usage. This compression storage method is efficient for both data storage and reading. At the start of iteration, all $\Gamma$s are stored in the CPU memory because they can exceed the maximum size of GPU memory. During iteration, the $\Gamma$s are copied into GPU memory in batches to update the new MFPs. Transport coefficients are then calculated using the updated MFPs. Next, the convergence is checked to determine if the loop should end. As a result, GPU memory limitations can be alleviated. Additionally, when CPU memory capacity is insufficient for storing $\Gamma$s, hard disks can be a backup option.

\section{Performance}\label{Sec:Perf}
We implemented our program on a CPU-GPU heterogeneous computer to compute the transport properties of representative metals and semiconductors. Next, we compared the results and running time of our program with those obtained from the CPU algorithm. Our program was executed on Nvidia Tesla v100 GPUs with 32 gigabyte memory, while two nodes of Intel Xeon Platinum 8260 CPUs with a total of 96 threads were used to run the CPU programs. Each physical core of the CPUs has two threads, with a clock frequency of 2.4 GHz. It is worth noting that the price of CPU is similar to that of the GPU.

To carry out the density functional theory and density functional perturbation theory calculations, we used the Quantum Espresso package. The local-density approximation (LDA) was employed to describe the exchange and correlation energy, and we utilized the EPW code to calculate the el-ph matrix elements in the Wannier basis.

Furthermore, we also performed calculations using two of the most widely used CPU programs: EPW (version 5.4.1) and Perturbo (version 2.0.0), along with Quantum Espresso 7.0.

\subsection{The convergence and precision of results}

\begin{figure*}[htb!]
	\includegraphics[width=13.5cm]{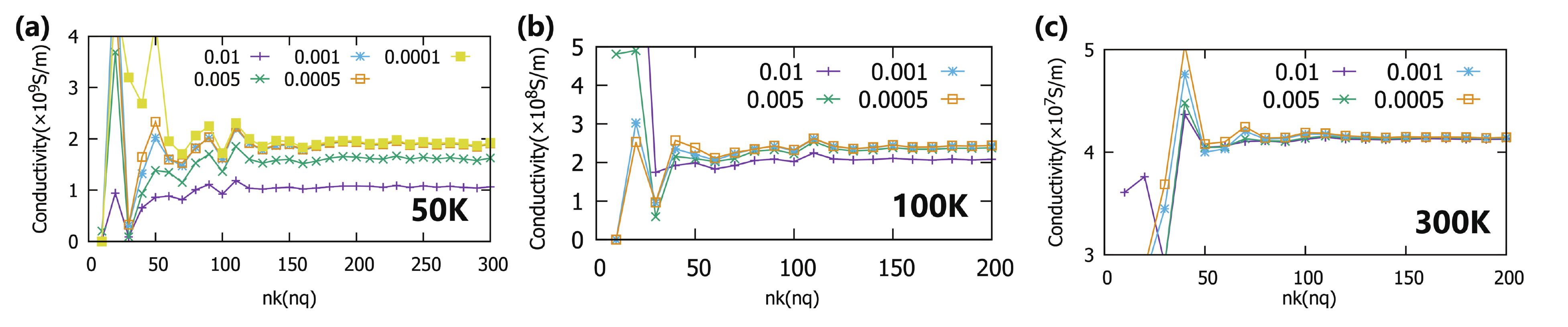}
	\caption{The convergence of the conductivity of aluminum at (a) 50 K, (b) 100 K, and (c) 300 K, with respect to the number of k and q points in each direction. The different colors indicate broadenings of Gaussian function. }
	\label{fig:al_meshconv}
\end{figure*}

At first, we calculated the conductivities of aluminum at different temperatures. Fig.\ref{fig:al_meshconv} shows the convergence of conductivity concerning interpolated $k$ and $q$ fine mesh. As discussed in Section \ref{intro}, the convergence is slower at lower temperatures. Additionally, the converged broadening $\eta$ at lower temperatures is smaller, requiring careful testing. The calculated conductivities at temperatures of 50 K, 100 K, and 300 K are $1.90 \times 10^9$ S/m, $2.41\times 10^8$ S/m, $4.14\times 10^7$ S/m, respectively. These results are in good agreement with the experimental data (being $2.12\times 10^9 $ S/m, $2.27\times 10^8 $ S/m, and $3.66\times 10^7 $ S/m\cite{cook1976thermal}).

\begin{figure*}[htb!]
	\includegraphics[width=13.5cm]{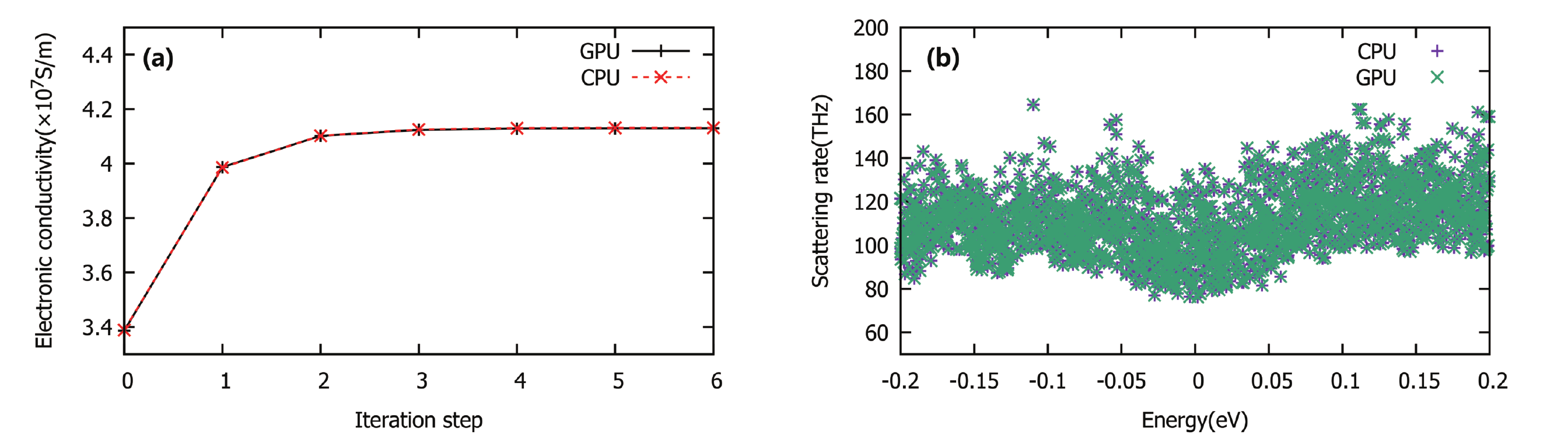}
	\caption{The comparison of (a) iteration process of the conductivity and (b) scattering rates of aluminum at 300 K by CPU and GPU.}
	\label{fig:al_sr_iter}
\end{figure*}

To verify the accuracy of the GPU algorithm, we compared the results obtained from the CPU and GPU algorithms. This was accomplished through the computation of the room temperature conductivity of aluminum using $100^3$ $k$ and $q$ sampling grids.
As is displayed in Fig.\ref{fig:al_sr_iter}, the differences in the scattering rates, iteration processing, and final results between the algorithms were all negligible. Hence, we can confirm that the GPU algorithm has been implemented correctly.

\begin{figure*}[htb!]
	\includegraphics[width=13.5cm]{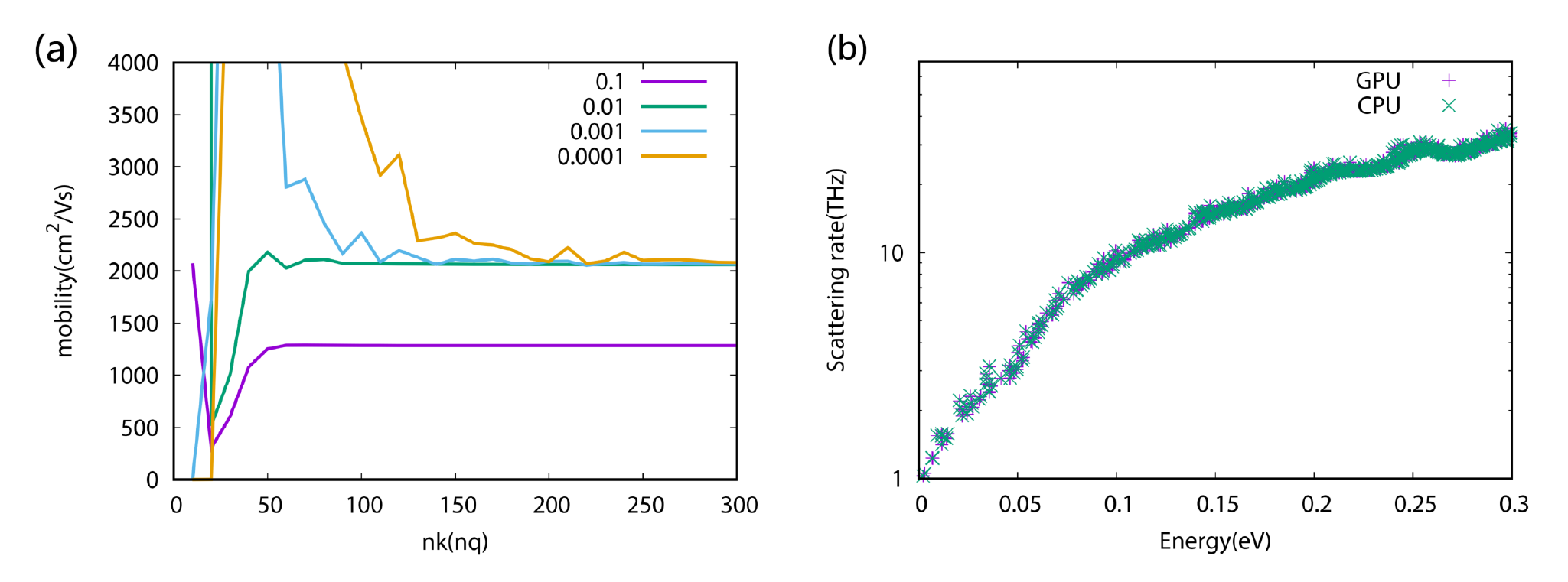}
	\caption{(a) The convergence of the mobility of silicon at 300 K, with respect to the density of k and q grids. Different colors indicate broadenings of Gaussian function. (b) The scattering rates of electron in silicon at 300 K. }
	\label{fig:si_sr}
\end{figure*}

We also compared the CPU and GPU algorithms with regards to the mobilities of silicon at room temperature. The results are shown in Fig.\ref{fig:si_sr}. Just like what we did on aluminum, we fully achieved the convergence and plotted it in Fig.\ref{fig:si_sr}(a). In Fig.\ref{fig:si_sr}(b), the consistency between CPU and GPU algorithm are verified. The converged results of the electronic mobility of silicon at 300 K is 2074 $\text{cm}^2\text{V}^{-1}\text{s}^{-1}$, which agrees with other theoretical and experimental results\cite{Canali_PRB75, logan1960impurity, norton1973impurity, li2015electrical, fiorentini2016thermoelectric,  PhysRevB.97.121201}. As the mobility does not change with the iteration, the iteration process of silicon is not shown.

In addition to aluminum and silicon, we have also tested on many systems and have found no inconsistencies.

\subsection{Speedup and detailed analysis}\label{result:time}
\begin{center}
    \begin{table*}[htb]
	\centering
	\caption{The comparison of performance of CPU and GPU program on different cases. Numbers in parentheses denote the sampling grids for both electron and phonon. }
    \begin{tabularx}{1.0\textwidth} { 
         >{\centering\arraybackslash}X
         >{\centering\arraybackslash}X 
         >{\centering\arraybackslash}X 
         >{\centering\arraybackslash}X  }
    \hline
             &  CPU (Mins)    & GPU (Mins) & speedup \\ 
    \hline
    Si ($200^3$)  &  163    &    5.5   &     30  \\
    Al ($200^3$)  &  4509   &    26    &     173 \\ 
    Be ($150^3$)  &  1627   &    61    &     27  \\
    As ($150^3$)  &  10397  &    65    &     160 \\
    Mo ($120^3$)  &  1858   &    53    &     35  \\
    \hline
    \end{tabularx}
	\label{Tab:timecomp_speedup}
\end{table*}
\end{center}

Achieving the high levels of precision required for this computation is so computationally intensive that it's challenging for a single CPU node to handle. 
However, the use of a heterogeneous computer with a GPU running the GPU-adapted algorithm can complete each task in a matter of hours. 
In Table.\ref{Tab:timecomp_speedup}, we compare the time consumption of the CPU and GPU algorithms and present the speedups in several cases. We observed that the speedup of the GPU algorithm ranged from 27 to 173 times when compared with the parallelized CPU implementation. 
Given the almost same price of the CPU and GPU resources, this magnitude of speedup is rather impressive. 

\begin{center}
    \begin{table*}[htb]
	\centering
	\caption{The comparison of the time cost of CPU and GPU program of each part.}
    \begin{tabularx}{1.0\textwidth} { 
         >{\centering\arraybackslash}X
         >{\centering\arraybackslash}X
         >{\centering\arraybackslash}X 
         >{\centering\arraybackslash}X 
         >{\centering\arraybackslash}X  }
    \hline
                 &           &  CPU (secs) & GPU (secs) & speedup \\ 
    \hline
    Prep + Iter  &           &  203       &    33    &      6.15  \\
    EPWI         & Dyn + Ham &  68200     &    1154   &     59.1 \\ 
                 & eph       &  167265    &    229    &     730 \\
                 & SR        &  3811      &    104    &     36.6  \\
    \hline
    \end{tabularx}
	\label{Tab:detailed_speedup}
\end{table*}
\end{center}

Since the algorithm is completely refactored, the comparison of the time cost of each function between CPU and GPU algorithm is no longer possible. Instead, we can only compare the time cost of each part. Table.\ref{Tab:detailed_speedup} displays the time cost of four parts of the code. 
The first one is the sum of the preprocessings and iterations, which contains several complex targets and necessitated writing most of the kernel functions ourselves. 
The second part comprises the preparation and diagonalization of the electron Hamiltonian and phonon dynamical matrix. The diagonalization of the matrix incurs the most significant computational cost, and we utilized the cuSolver library to handle it. 
The third part is the el-ph matrix element interpolation, which involves matrix multiplications and is handled by the cuBLAS library. 
The last part, SR, computes scattering rates and $\Gamma$ in the EPWI loops. Same as the first part, this part was also computed using kernel functions we authored.

The third part, which involves matrix multiplication, experiences the greatest boost from GPU acceleration since the GPU performs best when accelerating matrix multiplication. In the second part, which involves matrix diagonalization, the GPU exhibits the second-best acceleration. The remaining parts involve piecemeal tasks that are not as easy to parallelize. These results are consistent with what we know about GPU characteristics

\subsection{The effects of batch method and batch size}\label{result:perf}
\begin{center}
	\begin{table*}[htb]
	\centering
    \caption{The detailed performance comparison between the two batching methods. $N_{\text{call}}$ denotes the number of calls to each of the batched kernel functions. $N_{\text{scat}}$ denotes the total number of scattering processes. In this case, $N_{\text{scat}}=3.4\times 10^7$. $t_{\text{EPWI}}$ is the time cost of EPWI processes while $t_{\text{Ham}}$ means the time cost of electronic Hamiltonian initialization and diagonalization. }
	\label{Tab:batchmode}
    \begin{tabularx}{1.0\textwidth} { 
         >{\centering\arraybackslash}X 
         >{\centering\arraybackslash}X 
         >{\centering\arraybackslash}X  }
     \hline
                & q-batch & k-batch \\ 
     \hline
     $N_{\text{call}}$ & $7.8\times 10^3$ & $9.4\times 10^5$ \\ 
     $N_{\text{scat}}/N_{\text{call}}$ & $4.4\times 10^3$ & $36.2$ \\ 
     $t_{\text{EPWI}}$ & $23$s & $1500$s \\ 
     $t_{\text{EPWI}}/N_{\text{scat}}$ & $6.4\times 10^{-7}$s & $4.1\times 10^{-5}$s \\ 
     $t_{\text{Ham}}$ & $7.23$s & $160$s \\ 
     $t_{\text{Ham}}/N_{\text{scat}}$ & $2\times 10^{-7}$s & $5\times 10^{-6}$s \\ 
	 GPU Occupation & 90\% & 75\% \\
     \hline
    \end{tabularx}
\end{table*}
\end{center}
The GPU algorithm utilizes batch processing in which computational tasks are organized into batches, and then batches of tasks are sequentially carried out. The maximum number of $(k,q)$ pairs in each batch is governed by the variable batch\_size, which is set to 5000 if not specified. During each iteration, a series of kernel functions are executed on the GPU to implement the batched calculation.

As discussed in Section \ref{algo:loop}, the transition from a k-batch algorithm to a q-batch algorithm significantly enhances performance. To provide further clarity, we conducted performance tests on aluminum using the k-batch method. The corresponding results are summarized in Table \ref{Tab:batchmode}. The total time cost increased nearly 50 times after transitioning from a q-batch method to a k-batch method.

The variable batch\_size constrains the maximum number of $(k,q)$ pairs that a single kernel function can handle in one call. Ideally, the number of calls to each kernel function would be $N_{\text{scat}}/batch\_size$, which is $3.4\times 10^7/5000=6800$ in the most efficient scenario. Thus, $N_{\text{scat}}/N_{\text{call}}$ can be considered as the number of $(k,q)$ pairs in each batch in this example.

As expected, the number of kernel function calls significantly increases from $7800$ to $9.4\times 10^5$ when transitioning from the q-batch to k-batch method. As the total number of processes are $3.4\times 10^7$ and remains constant regardless of the method, the average number of processes per call decreases from 5000 to 40. 
Since 5000 is the maximum number of processes in each batch, that means the q-batch method maximizes the benefits of batch computing. On the contrary, the k-batch method underutilizes the capacity of the device leading to wasted resources. 

\begin{figure*}[!htb]
	\includegraphics[width=13.5cm]{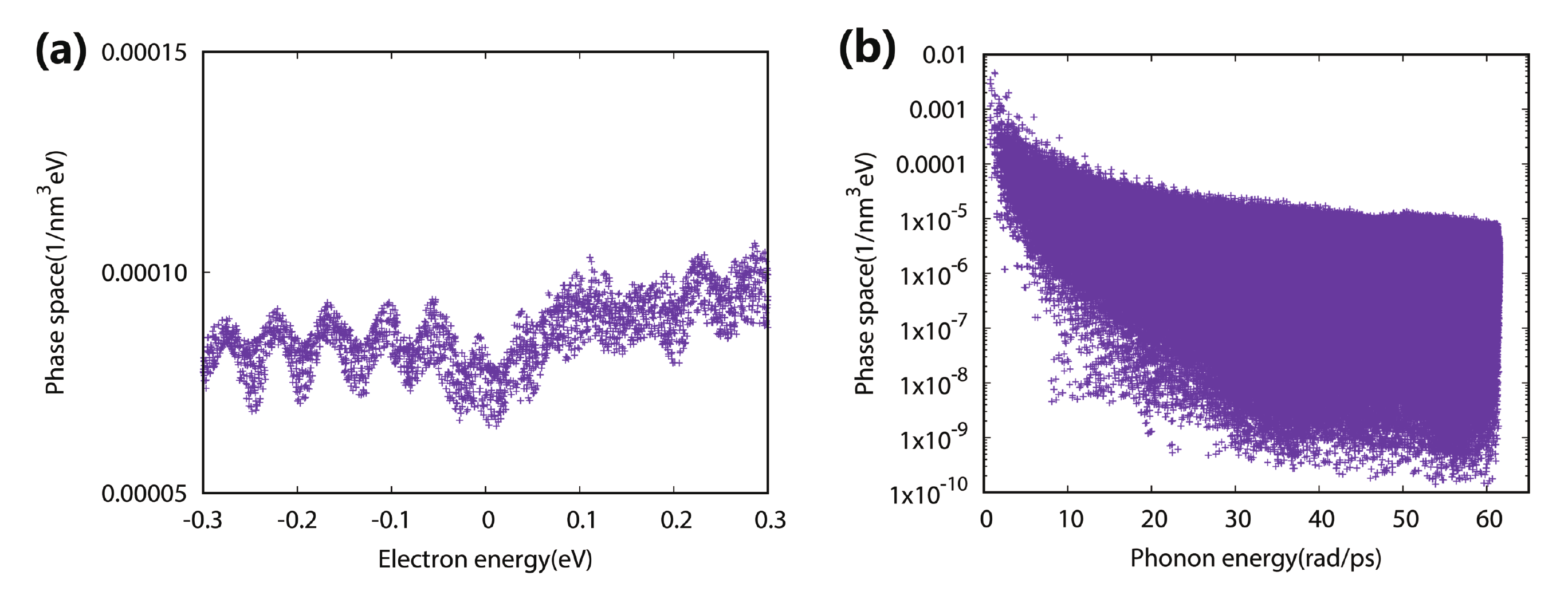}
	\caption{The volume of phase space of relevant (a) electron and (b) phonon states in aluminum with $100^3$ k and q mesh. Effective phonon states are much more than electron, and the phase space of most phonons are far small than electron.}
	\label{fig:Al_phase}
\end{figure*}

The phase space of effective electron states and phonon states was calculated, and the results are shown in Fig.\ref{fig:Al_phase}. 
The number of effective electron states is approximately $5\times 10^4$, whereas almost all phonon wavevectors are valid. 
As expected, we also observed that the phase space of most phonons is significantly smaller than that of electrons. As a result, the q-batch method is capable of accommodating larger batches.

\begin{figure}[htb!]
	\includegraphics[width=7cm]{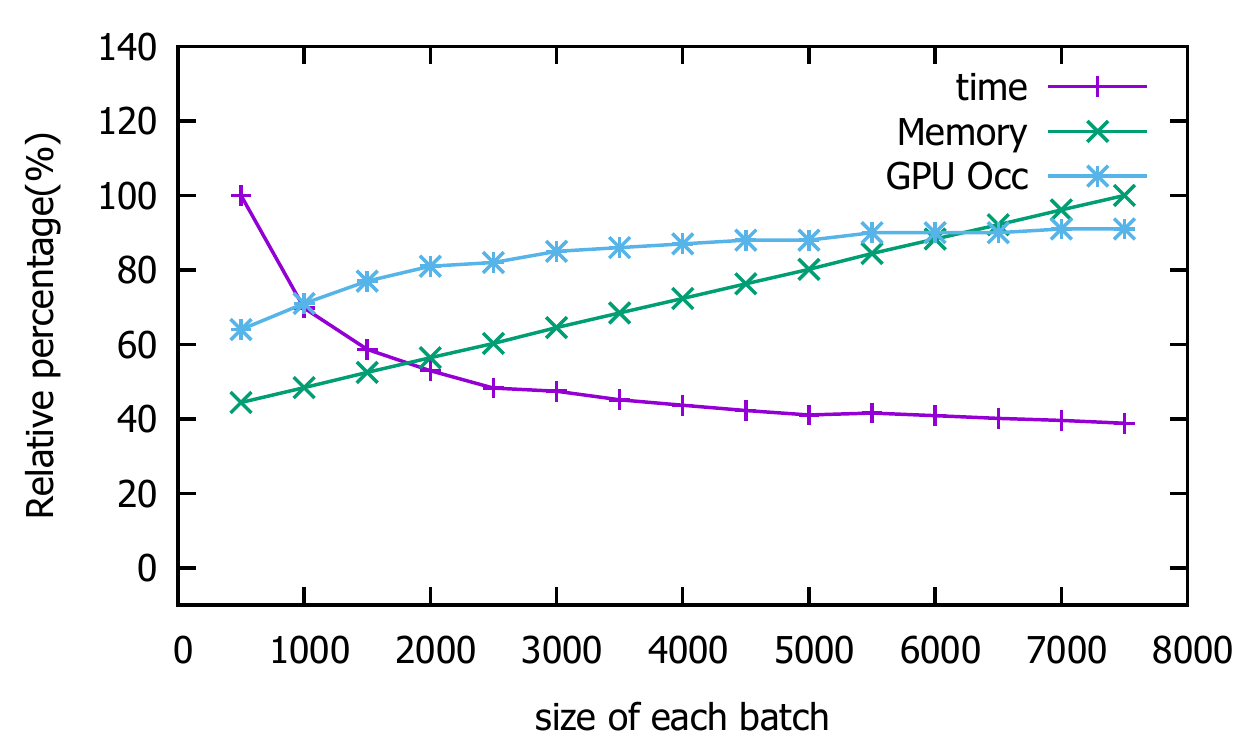}
	\caption{The relative time cost, memory usage, and GPU occupation versus batch\_size. We treat the time cost when the batch\_size=500, total GPU memory and GPU fully occupation as 100\%, respectively. The GPU memory runs out when the batch\_size is larger than 7500. }
	\label{fig:batchsize}
\end{figure}

The value of batch\_size is noteworthy too. A large batch\_size can increase the utilization rate of the GPU, but also increase the memory usage. Hence, a modest batch\_size is beneficial for calculation. In order to give quantitative results, we test the algorithm with a stepwise increase in the value of batch\_size from 500 to 7000. The relative time cost and device utilization are shown in Fig.\ref{fig:batchsize}. The memory usage grows linearly with the increase of batch\_size and used up at batch\_size=7500. While the performance reaches 90\% of its limit with batch\_size above 4000.  For a larger system, the average time cost for computing each process is much larger and the optimum batch\_size could be much smaller.

\subsection{ A brief comparison with other CPU programs }\label{result:pertEPW}
Two of the most widely-used programs, Perturbo and EPW, were tested and compared with the GPU implementation. The results are presented below.


\begin{figure*}[htb!]
	\includegraphics[width=13.5cm]{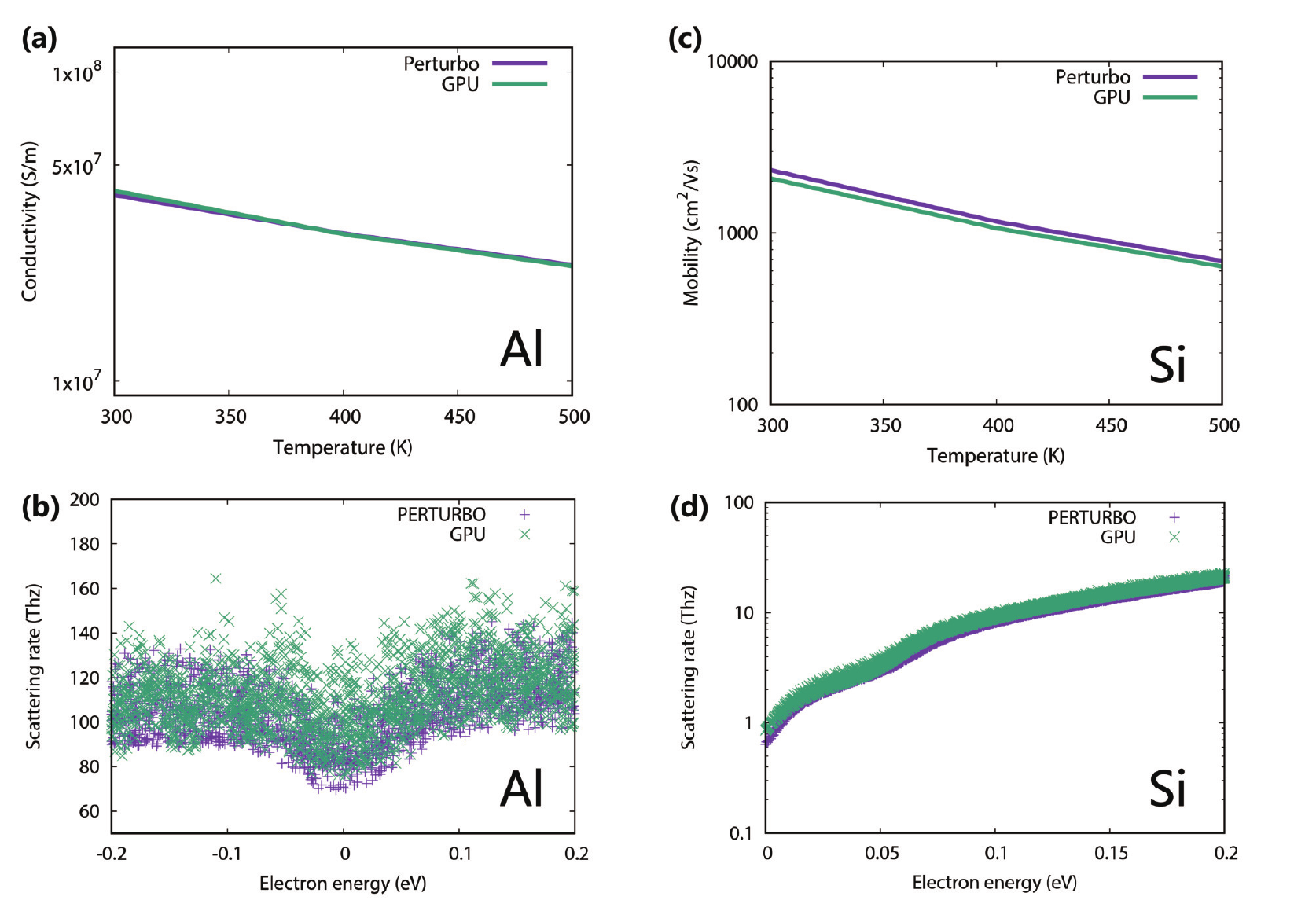}
	\caption{The results by comparing the GPU algorithm with Perturbo code (a CPU-only program) on the (a) conductivity and (b) scattering rates of aluminum, as well as the (b) mobility and (d) scattering rates of silicon. }
	\label{fig:Pert}
\end{figure*}

The mobility of silicon and the conductivity of aluminum with $200^3$ k and q fine mesh are also calculated by Perturbo code. They are compared with GPU and the results are shown in Fig.\ref{fig:Pert}. As shown in Fig.\ref{fig:Pert}(a), from 300 K to 500 K, the difference on the conductivity of aluminum are negligible. Additionally, the scattering rates of aluminum at 300 K (shown in Fig.\ref{fig:Pert}(b)) are comparable to our results. Similarly, the mobilities (Fig.\ref{fig:Pert}(c)) of silicon between 300 K and 500 K and the scattering rates (Fig.\ref{fig:Pert}(d)) of silicon at 300 K and are also consistent with our results. Notably, the Perturbo calculation for silicon consumed more than 80 times the time required for our GPU implementation, which is a remarkable acceleration. Incorporating iteration processes into the computation of aluminum, we achieved a speedup of over 200 using the GPU algorithm. However, to make a fair comparison, we should assess the computational time for scattering rates since Perturbo recalculates and saves the el-ph matrix elements to disk during the iteration process. In contrast, the GPU algorithm already calculates and stores the elements in memory. Accordingly, we achieved a speedup of 120 for the calculation of scattering rates. However, owing to various differences in implementation details, there are still small differences in the results.

The mobility of silicon and the conductivity of aluminum are also calculated using the EPW code. However, the implementation of EPW differs significantly, which greatly diminishes the significance of comparison.  
The strategy employed by EPW to select valid scattering processes is more cautious and considerably more expensive. It determines the validity of scattering processes based on the transition rate $\Gamma$, whereas we determine them solely based on the law of conservation of energy. 
The approach employed by EPW delays the sieving process and requires the calculation of additional electron-phonon matrix elements. 

Nevertheless, despite the limitations of making a direct comparison between EPW and the GPU method, we still present the obtained results. Using a grid size of $200^3$, the EPW implementation for silicon requires 55 times more time compared to the GPU code. Similarly, when using a grid size of $100^3$ and no iteration for aluminum, the EPW approach takes 600 times longer than the GPU code. The calculated mobility of silicon and conductivity of aluminum are 1652 $\text{cm}^2\text{V}^{-1}\text{s}^{-1}$ and $3.63\times 10^7$ S/m, respectively.


Among other programs with similar objectives, the PHOEBE code\cite{cepellotti2022phoebe} is the one worth mentioning, as it accelerates the calculation with GPUs. However, the reported speedup of PHOEBE, which is only 3, is noticeably slower than our technique. This can be attributed to their use of the Kokkos library, a hardware-agnostic programming paradigm. Evidently, compared to our fully GPU-adapted algorithm, hardware-agnostic models are incapable of fully exploiting the potential of GPUs.

Thus, we can conclude that GPU algorithms improve performance by no less than two orders of magnitude when compared to the commonly-used similar CPU programs.

\section{Summary}\label{Sec:Conclusion}
In this paper, we implemented a high-performance GPU algorithm for the calculation of intrinsic transport properties based on electron-phonon Wannier interpolation. With comparable financial costs for CPU and GPU, our GPU implementation can accelerate the EPWI+BTE method by 30 to 170 times, against both our benchmark CPU code and existing ones. 
Firstly, we showed the structure of our GPU algorithm, discussed the main construction for parallelization that is crucial for achieving high performance. Then we calculated the transport properties of a few typical materials by the GPU algorithm and compared the results with CPU benchmark results. Based on these data, we confirmed the accuracy and the efficiency of the GPU algorithm. Finally, we analyzed the reason for such a speedup ratio and the effects of batching method and batch size. 

The proposed algorithm significantly eases the computation of intrinsic mobilities and conductivity for materials. This benefit is particularly helpful for obtaining fully converged results at low temperatures and computing properties of relatively complex systems. 
Moreover, given the extensive impact of EPIs and the variety of applications of the EPWI method, the algorithm can also be used to study other EPI-related properties with some moderate modifications. 

\section{Acknowledgment}\label{Ackn}
We acknowledges support from the Natural Science Foundation of China (NSFC) (Grant No. 12174261), and the GuangDong Basic and Applied Basic Research Foundation (Grants No. 2021A1515010042 and No. 2023A1515010365). 
We also thank Xiaoguang Li for providing the GPU supercomputer as a development and testing platform.

\bibliographystyle{apsrev4-2}
\bibliography{Citations}

\end{document}